\documentstyle[amssymb,latexsym,psfig]{ag}

\title[Galaxy Collisions]{Starbirth in Galaxy Collisions}

\author[R.  de Grijs]{Richard de Grijs\\
Department of Physics \& Astronomy, The University of Sheffield, Hicks
Building, Hounsfield Road, Sheffield S3 7RH, UK;\\ Email:
R.deGrijs@sheffield.ac.uk}

\pubyear{2003}

\begin{document}
\maketitle

\begin{abstract}
Young, massive star clusters are the most notable and significant end products
of violent star-forming episodes triggered by galaxy collisions, mergers, and
close encounters. Their contribution to the total luminosity induced by such
extreme conditions dominates, by far, the overall energy output due to the
gravitationally-induced star formation. The general characteristics of
these newly-formed clusters (such as their masses, luminosities, and sizes)
suggest that at least a fraction may eventually evolve into equal, or
perhaps slightly more massive, counterparts of the abundant old globular
cluster systems in the local Universe. Establishing whether or not such an
evolutionary connection exists requires our detailed knowledge of not only the
physics underlying the evolution of ``simple'' stellar populations, but also
that of cluster disruption in the time-dependent gravitational potentials of
interacting galaxies. Initial results seem to indicate that proto-globular
clusters do indeed continue to form today, which would support hierarchical
galaxy formation scenarios.
\vspace*{1cm}{~}

\end{abstract}

\section{Extreme environmental conditions}

Stars rarely form in isolation. In fact, star formation in galaxies generally
occurs in extended regions, where the fragmentation of the giant molecular
clouds (GMCs) making up a significant fraction of a galaxy's interstellar
medium (ISM) leads to the (almost simultaneous) gravitational collapse of
multiple GMC subclumps. It is well known that the vast majority of stars in
the Milky Way, and in nearby galaxies out to distances where individual stars
and a variety of star cluster-type objects can be resolved by high-resolution
observations, are found in groups ranging from binary stars to ``OB'' or ``T
Tauri'' associations (young star-forming regions dominated by a small number
of massive stars), open cluster-type objects, compact, old ``globular'' and
young massive clusters, to supermassive clusters often confusingly referred to
as ``super star clusters''. The nearest examples of these latter objects
include the Milky Way star-forming region NGC 3603, and the giant starburst
region 30 Doradus with its central star cluster R136 in the Large Magellanic
Cloud.

In addition to a fraction of the more massive unbound OB associations, our
Milky Way galaxy contains two main populations of gravitationally bound
clusters with masses exceeding $\sim 10^3 {\rm M}_\odot$. The Milky Way's
globular cluster population, consisting of some 150 compact objects with a
median mass of $M_{\rm cl} \simeq 3 \times 10^5 {\rm M}_\odot$ (e.g., Harris
1996, 2001), is predominantly old, with ages $\gtrsim 8-10$ billion years. The
much larger open cluster population (with a likely Galactic total number $\sim
10^5$), on the other hand, is dominated by significantly younger ages
(although open clusters up to the lower age limit of the globular cluster
population do exist) and lower masses ($10-10^4 {\rm M}_\odot$). Although the
older open clusters are undoubtedly gravitationally bound objects, their lower
masses and more diffuse structures make them much more vulnerable to disk (and
bulge) shocking when they pass through the Milky Way disk (or close to the
bulge) on their orbits, thus leading to enhanced cluster evaporation. These
objects are therefore unlikely globular cluster progenitors. It appears that
the conditions for the formation of compact, massive star clusters -- that
have the potential to eventually evolve into globular cluster-type objects by
the time they reach a similar age -- are currently not present in the Milky
Way, or at best to a very limited extent.


The production of luminous, massive yet compact star clusters seems to be a
key feature of the most intense star-forming episodes. Such so-called
``starbursts'' normally occur at least once during the lifetimes of the vast
majority of galaxies. The defining properties of young massive star clusters
(with masses often significantly in excess of $M_{\rm cl} = 10^5 {\rm
M}_\odot$, i.e., the median mass of the abundant old globular clusters in the
local Universe) have been explored in intense starburst regions in several
dozen galaxies, often involved in gravitational interactions of some sort
(e.g., Holtzman et al. 1992, Whitmore et al. 1993, O'Connell et al. 1994,
Conti et al. 1996, Watson et al. 1996, Carlson et al. 1998, de Grijs et al.
2001, 2003a,b,c,d,e).


An increasingly large body of observational evidence suggests that a large
fraction of the star formation in starbursts actually takes place in the form
of such concentrated clusters, rather than in small-scale star-forming
``pockets''. Young massive star clusters are therefore important as
benchmarks of cluster formation and evolution. They are also important as
tracers of the history of star formation of their host galaxies, their
chemical evolution, the initial mass function (IMF; i.e., the proportion of
low to high-mass stars at the time of star formation), and other physical
characteristics in starbursts.

In a detailed study of the young star cluster population associated with the
fading starburst region ``B'' in the prototype nearby starburst galaxy M82 (de
Grijs et al. 2001), we concluded that the last tidal encounter between M82 and
its large neighbour spiral galaxy M81, which occurred about $500 - 800$ Myr
ago (Brouillet et al. 1991, Yun 1999) had a major impact on what was probably
an otherwise normal, quiescent, disk galaxy. It caused a concentrated
starburst, resulting in a pronounced peak in the clusters' age distribution,
roughly 1 billion years ago (de Grijs et al. 2001, 2003c, Parmentier et al.
2003). The enhanced cluster formation decreased rapidly within a few hundred
Myr of its peak. However, general (``field'') star formation activity
continued in the galactic disk of M82's region B, probably at a much lower
rate, until $\sim 20$ Myr ago.


The evidence for the decoupling between cluster and field star formation
is consistent with the view that star cluster formation requires special
conditions, such as large-scale gas flows, in addition to the presence
of dense gas (cf.  Ashman \& Zepf 1992, Elmegreen \& Efremov 1997). 
Such conditions occur naturally in the extreme environments of colliding
or otherwise gravitationally interacting galaxies. 

Using optical observations of the ``Mice'' and ``Tadpole'' interacting
galaxies (NGC 4676 and UGC 10214, respectively; see Fig. 1) -- based on a
subset of the Early Release Observations obtained with the {\sl Advanced
Camera for Surveys} on board the {\sl Hubble Space Telescope} -- and the novel
technique of pixel-by-pixel analysis of their colour-colour and
colour-magnitude diagrammes, we deduced the systems' star and star cluster
formation histories (de Grijs et al. 2003e).

\begin{figure*}
\psfig{file=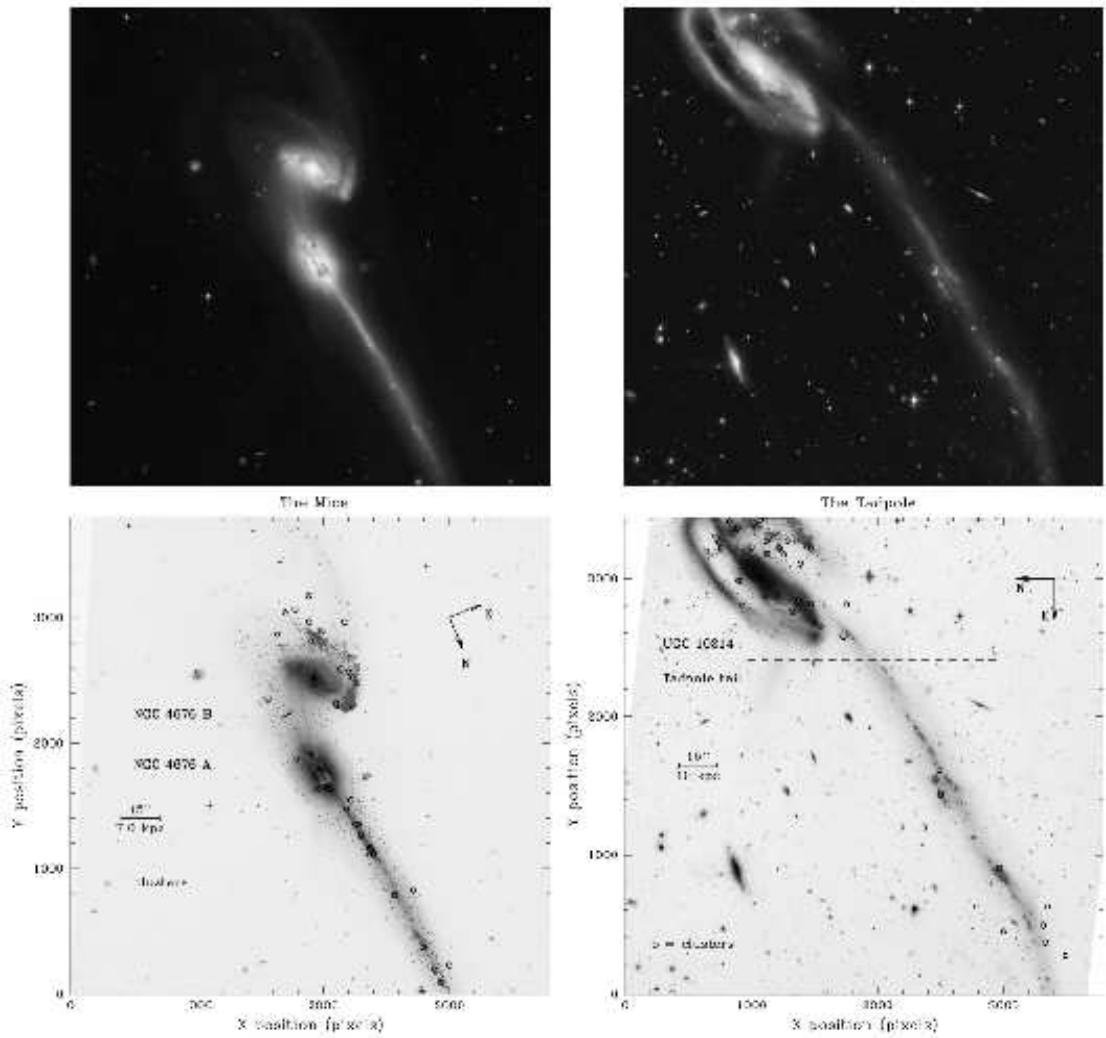,width=17cm}
\caption[]{{\it top:} {\sl Hubble Space Telescope} images of the Mice {\it
(left)} and the Tadpole {\it (right)} interacting systems. Source: press
release STScI-2002-11; {\it bottom:} Locations of the actively star-forming
regions in both systems, overlaid on grey-scale versions of the {\sl Hubble
Space Telescope} images. The locations of the star cluster candidates are
indicated by circles.}
\end{figure*}

In both of these interacting systems we find several dozen bright (and
therefore massive) young star clusters (or, alternatively, compact star
forming regions), which overlap spatially with regions of active star
formation in the galaxies' tidal tails and spiral arms (from a comparison with
H$\alpha$ observations that trace active star formation; Hibbard \& van Gorkom
1996). We estimate that the main gravitational interactions responsible for
the formation of these clusters occurred $\sim 150 - 200$ Myr ago.

In fact, we show that star cluster formation is a major mode of star formation
in galactic interactions, with $\gtrsim 35$\% of the active star formation in
encounters occurring in star clusters (de Grijs et al. 2003e). The tidal tail
of the Tadpole system is dominated by star forming regions, which occupy some
60\% of the total area covered by the tail; they contribute $\sim 70$\% of the
total flux in our bluest filter (decreasing to $\sim 40$\% in the reddest
filter; see Fig. 1, bottom row). If the encounter occurs between unevenly
matched, gas-rich galaxies then, as expected, the effects of the gravitational
interaction are much more pronounced in the smaller galaxy. For instance, when
we compare the impact of the interaction as evidenced by star cluster
formation between M82 (de Grijs et al. 2001, 2003b,c) and M81 (Chandar et al.
2001), or the star cluster formation history in the ``Whirlpool Galaxy'' M51
(Bik et al. 2003), which is currently in the process of merging with the
smaller spiral galaxy NGC 5194, the evidence for enhanced cluster formation in
the larger galaxy is minimal if at all detectable.

The NGC 6745 system (the ``Bird's Head Galaxy''; Fig.  2) represents a
remarkably violently star-forming interacting pair of unevenly matched
galaxies.  The optical morphology of NGC 6745, and in particular the
locations of the numerous bright blue star-forming complexes and compact
cluster candidates, suggest a recent tidal passage by the small northern
companion galaxy (NGC 6745c; nomenclature from Karachentsev et al. 
1978) across the eastern edge of the main galaxy, NGC 6745a.  The high
relative velocities of the two colliding galaxies likely caused ``ram
pressure'' at the surface of contact between both galaxies, which -- in
turn -- is responsible for the triggering of enhanced star and cluster
formation, most notably in the interaction zone in between the two
galaxies, NGC 6745b (cf.  de Grijs et al.  2003a).  The smaller galaxy,
however, does not show any significant enhanced cluster formation, which
is most likely an indication that it contains very little gas.

\begin{figure}
\psfig{figure=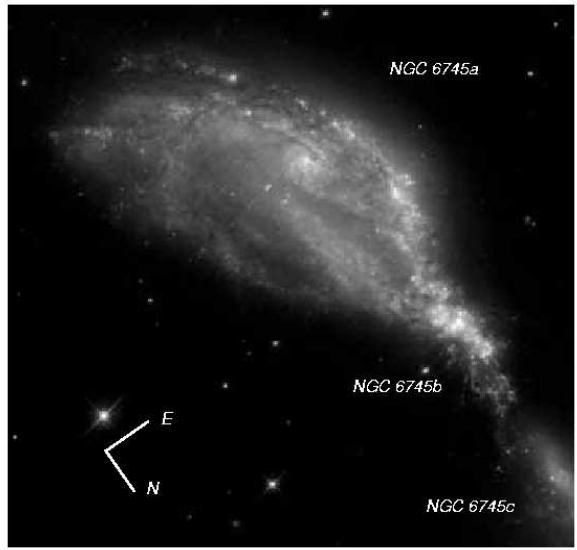,width=8.5cm}
\caption[]{{\sl Hubble Space Telescope} image of the NGC 6745 interacting
system, the ``Bird's Head'' galaxy. The tiny northern companion galaxy NGC
6745c has likely caused the havoc seen in the form of active, blue star
(cluster) formation in NGC 6745b and along the eastern spiral-arm-like feature
of the main galaxy component.}
\end{figure}

For the NGC 6745 young cluster system we derive a median age of $\sim 10$ Myr.
Based on the age distribution of the star clusters, and on the H{\sc i}
morphology of the interacting system (from high-resolution observations with
the Very Large Array), we confirm the interaction-induced enhanced cluster
formation scenario once again. NGC 6745 contains a significant population of
high-mass clusters, with masses in the range $6.5 \lesssim \log( M_{\rm
cl}/{\rm M}_\odot ) \lesssim 8.0$. These clusters do not have counterparts
among the Milky Way globular clusters (e.g., Mandushev et al. 1991, Pryor \&
Meylan 1993), but are similar to or exceed the spectroscopically confirmed
mass estimates of the ``super star clusters'' (SSCs) in M82 (e.g., M82 F and
L; see Smith \& Gallagher 2001) and the Antennae galaxies (Mengel et al.
2002). We caution, however, that these massive SSC candidates may not be
gravitationally bound objects, but more diffuse star forming regions or
aggregates of multiple unresolved clusters instead. Nevertheless, we measure a
very compact ``effective'' radius (i.e., the radius within which half of an
object's light is contained) for the most massive object ($M_{\rm cl} \simeq
5.9 \times 10^8 {\rm M}_\odot$) of only $R_{\rm eff} \sim 16$ pc. However,
this object appears very elongated, or may in fact be a double cluster. We
should also keep in mind that this high mass estimate is a strong function of
the (low) metallicity assumed; if we had assumed (higher) solar metallicity
for this object, the derived age would have been significantly smaller ($\sim
10-20$ Myr vs. $\sim 1$ Gyr assumed in our study), and the mass could be
smaller by a factor of $\ga 10$. Even so, if we could confirm this mass
estimate spectroscopically, either of the subcomponents would be the most
massive cluster known to date, significantly exceeding cluster W3 in NGC 7252,
which has a mass of about $(3-18) \times 10^7 {\rm M}_\odot$, depending on the
age, metallicity and IMF assumed (Schweizer \& Seitzer 1998; Maraston et al.
2001). Our detection of such massive SSCs in NGC 6745, which are mostly
located in the intense interaction zone, supports the scenario that such
objects form preferentially in the extreme environments of interacting
galaxies.

\section{An evolutionary connection?} 

The (statistical) derivation of galaxy formation and evolution scenarios using
their star cluster systems as tracers is limited to the study of integrated
cluster properties (such as their luminosities, sizes, masses, ages and
metallicities) for all but the nearest galaxies, even at {\sl Hubble Space
Telescope} spatial resolution.

The question remains, therefore, whether or not at least a fraction of the
young compact star clusters seen in abundance in extragalactic starbursts, are
potentially the progenitors of globular cluster-type objects in their host
galaxies. If we could settle this issue convincingly, one way or the other,
the implications of such a result would have profound and far-reaching
implications for a wide range of astrophysical questions, including (but not
limited to) our understanding of the process of galaxy formation and assembly,
and the process and conditions required for star (cluster) formation. Because
of the lack of a statistically significant sample of similar nearby objects,
however, we need to resort to either statistical arguments or to the
painstaking approach of one-by-one studies of individual objects in more
distant galaxies, as outlined below. With the ever increasing number of
large-aperture ground-based telescopes equipped with state-of-the-art
high-resolution spectroscopic detectors and the wealth of observational data
provided by the {\sl Hubble Space Telescope} we may now be getting close to
resolving this important issue. It is of paramount importance, however, that
theoretical developements go hand in hand with observational advances.

The present state-of-the-art teaches us that the sizes, luminosities, and --
in several cases -- spectroscopic mass estimates of most (young) extragalactic
star cluster systems are fully consistent with the expected properties of
young Milky Way-type globular cluster progenitors (e.g., Meurer 1995, van den
Bergh 1995, Ho \& Filippenko 1996a,b, Schweizer \& Seitzer 1998, de Grijs et
al. 2001, 2003d). For instance, for the young massive star cluster system in
the centre of the nearby starburst spiral galaxy NGC 3310, we find a median
mass of $\langle \log(M_{\rm cl}/{\rm M}_\odot) \rangle = 5.24 \pm 0.05$ (de
Grijs et al. 2003d); their mass distribution is characterised by a Gaussian
width of $\sigma_{\rm Gauss} \simeq 0.33$ dex. In view of the uncertainties
introduced by the poorly known lower-mass slope of the stellar IMF ($m_\ast
\lesssim 0.5 {\rm M}_\odot$; see below), our median mass estimate of the NGC
3310 cluster system -- which was most likely formed in a (possibly extended)
global burst of cluster formation $\sim 3 \times 10^7$ yr ago -- is remarkably
close to that of the Milky Way globular cluster system (cf. de Grijs et al.
2003d; Fig. 3).

[Recent determinations of the stellar IMF deviate significantly from a (solar
neighbourhood-like) Salpeter-type IMF at low masses, in the sense that the
low-mass stellar IMF may well be significantly flatter than the Salpeter
slope. The implication of using a Salpeter-type IMF for our cluster mass
determinations is therefore that we may have {\it overestimated} the
individual cluster masses (although the relative mass distributions of our
cluster samples remain unaffected). Therefore, from a comparison with the more
modern IMF parametrisation of Kroupa, Tout \& Gilmore (1993) we derive that --
depending on the adopted slope for the lowest mass range, below $m_\ast = 0.08
{\rm M}_\odot$ -- we may have overestimated our individual cluster masses by a
factor of $\sim 1.5 - 2.4$.]

\begin{figure}
\psfig{figure=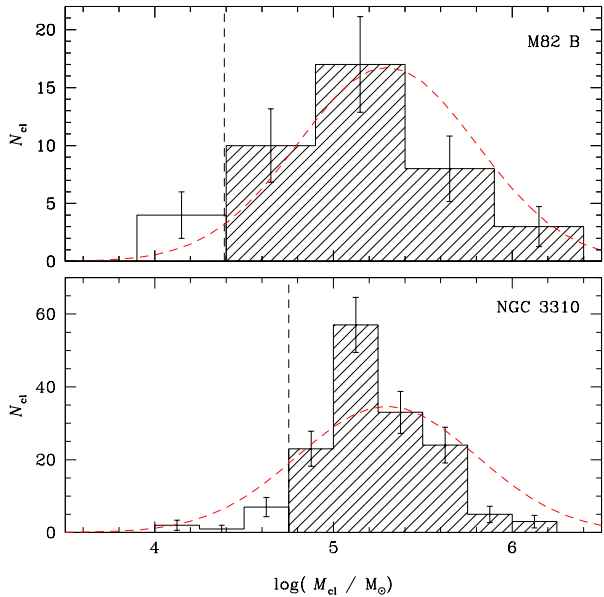,width=8.5cm}
\caption[]{Logarithmic mass distributions of the intermediate-age, $\sim 1$
billion year-old star clusters in M82 B {\it (top)}, and the young ($\sim 30$
Myr-old) cluster population dominating the central regions of the ringed and
barred starburst galaxy NGC 3310 {\it (bottom)}. For comparison, the red
dashed Gaussian distributions show the mass distribution of the Milky Way
globular cluster system, normalised to the same number of clusters in each
population. The vertical dashed lines indicate our 50 per cent observational
completeness limits, with severe incompleteness affecting the clusters
populating the open histograms.}
\end{figure}

However, the postulated evolutionary connection between the recently formed
massive star clusters in regions of violent star formation and starburst
galaxies, and old globular clusters similar to those in the Milky Way, the
Andromeda galaxy, the giant elliptical galaxy M87 at the centre of the Virgo
cluster, and other old elliptical galaxies is still a contentious issue. The
evolution and survivability of young clusters depend crucially on the stellar
IMF of their constituent stars (cf. Smith \& Gallagher 2001): if the IMF is
too shallow, i.e., if the clusters are significantly depleted in low-mass
stars compared to (for instance) the solar neighbourhood, they will disperse
within a few orbital periods around their host galaxy's centre, and likely
within about a billion years of their formation (e.g., Smith \& Gallagher
2001, Mengel et al. 2002).


Ideally, one would need to obtain (i) high-resolution spectroscopy (e.g., with
8m-class ground-based telescopes) of all clusters in a given cluster sample in
order to obtain dynamical mass estimates (we will assume, for the purpose of
the present discussion, that our young clusters can be approximated as systems
in full virial equilibrium, so that the widths of their absorption lines
reflect the clusters' internal velocity dispersions and therefore their
masses) and (ii) high-resolution imaging (e.g., with the {\sl Hubble Space
Telescope}) to measure their luminosities. One could then estimate the
mass-to-light (M/L) ratios for each cluster, and their ages from the features
in their spectra. The final, crucial analysis would involve a direct
comparison between the clusters' locations in the M/L ratio vs. age diagramme
with models of so-called ``simple stellar populations'' (i.e., stellar
populations of a single metallicity formed in an instantaneous burst of star
formation) governed by a variety of IMF descriptions (cf. Smith \& Gallagher
2001, Mengel et al. 2002).

However, individual young star cluster spectroscopy, feasible today with
8m-class telescopes for the nearest systems, is very time-consuming, since
observations of large numbers of clusters are required to obtain statistically
significant results. Instead, one of the most important and most widely used
diagnostics, both to infer the star (cluster) formation history of a given
galaxy, and to constrain scenarios for its expected future evolution, is the
distribution of cluster luminosities, or -- alternatively -- their associated
masses, commonly referred to as the cluster luminosity and mass functions
(CLF, CMF), respectively.

Starting with the seminal work by Elson \& Fall (1985) on the young cluster
system in the Large Magellanic Cloud (with ages $\lesssim 2 \times 10^9$ yr),
an ever increasing body of evidence, mostly obtained with the {\sl Hubble
Space Telescope}, seems to imply that the CLF of young star clusters (YSCs) is
well described by a power law of the form $N_{\rm YSC}(L) {\rm d} L \propto
L^{\alpha} {\rm d} L$, where $N_{\rm YSC}(L) {\rm d} L$ is the number of YSCs
with luminosities between {\it L} and $L + {\rm d} L$, and $-2 \lesssim \alpha
\lesssim -1.5$ (e.g., Elson \& Fall 1985, Elmegreen \& Efremov 1997, Whitmore
et al. 2002, Bik et al. 2003, see also Elmegreen 2002). On the other hand, for
the old globular cluster systems in the local Universe, with ages $\gtrsim 10$
billion years, the CLF shape is well established to be roughly Gaussian
(Whitmore et al. 1995, Harris 1996, 2001, Harris et al. 1998). This shape
(characterised by its peak -- or turn-over -- magnitude and width) is almost
universal, showing only a weak dependence on the metallicity and mass of the
host galaxy (e.g., Harris 1996, 2001, Whitmore et al. 2002).

This type of observational evidence has led to the popular -- but thus far
mostly speculative -- theoretical prediction that not only a power-law, but
{\it any} initial CLF (and CMF) will be rapidly transformed into a Gaussian
distribution because of (i) stellar evolutionary fading of the
lowest-luminosity (and therefore lowest-mass) clusters to below the detection
limit; and (ii) disruption of the low-mass clusters due both to interactions
with the gravitational field of the host galaxy, and to cluster-internal
two-body relaxation effects (such as caused by star-star collisions and the
resulting redistribution of mass inside the cluster) leading to enhanced
cluster evaporation (e.g., Elmegreen \& Efremov 1997, Gnedin \& Ostriker 1997,
Ostriker \& Gnedin 1997, Fall \& Zhang 2001).

We recently reported the first discovery of an approximately Gaussian CLF (and
CMF) for the star clusters in M82 B formed roughly simultaneously in a
pronounced burst of cluster formation (Fig. 3, top). This provides the very
first sufficiently deep CLF (and CMF) for a star cluster population at
intermediate age (of $\sim 1$ billion years), which thus serves as an
important benchmark for theories of the evolution of star cluster systems.

The shape of the CLF (CMF) of young cluster systems has recently attracted
renewed theoretical and observational attention. Various authors have pointed
out that for young star clusters exhibiting an age range, one must first
correct their CLF to a common age before a realistic assessment of their
evolution can be achieved (e.g., Meurer 1995, Fritze--v. Alvensleben 1998,
1999, de Grijs et al. 2001, 2003b,c). This is particularly important for young
cluster systems exhibiting an age spread that is a significant fraction of the
system's median age, because of the rapid evolution of the colours and
luminosities of star clusters at young ages (below $\sim 1$ Gyr). Whether the
observed power laws of the CLF and CMF for young star cluster systems are
intrinsic to the cluster population or artefacts due to the presence of an age
spread in the cluster population -- which might mask a differently shaped
underlying distribution -- is therefore a matter of debate (see, e.g., Carlson
et al. 1998, Zhang \& Fall 1999, Vesperini 2000, 2001).

Nevertheless, the CLF shape and characteristic luminosity of the M82 B cluster
system is nearly identical to that of the apparently universal CLFs of the old
globular cluster systems in the local Universe (e.g., Whitmore et al. 1995,
Harris 1996, 2001, Ashman \& Zepf 1998, Harris et al. 1998). This is likely to
remain virtually unchanged for a Hubble time (i.e., the present age of the
Universe, defined as $H_0^{-1}$, where $H_0$ is the Hubble constant), if the
currently most popular cluster disruption models hold. With the very short
characteristic cluster disruption time-scale governing M82 B (see below), its
cluster mass distribution will evolve towards a higher characteristic mass
scale than that of the Milky Way globular cluster system by the time it
reaches a similar age. Thus, this evidence, combined with the similar cluster
sizes (de Grijs et al. 2001), lends strong support to a scenario in which the
current M82 B cluster population will eventually evolve into a significantly
depleted old Milky Way-type globular cluster system dominated by a small
number of high-mass clusters. This implies that globular clusters, which were
once thought to be the oldest building blocks of galaxies, are still forming
today in galaxy interactions and mergers. However, they will likely be more
metal-rich than the present-day old globular cluster systems. This is simply
so because, at the formation epoch of the current generation of old globular
clusters, the metal content of their host galaxies' ISM was still close to
primordial, and therefore almost pristine; in the generally accepted scenario
of the origin of globular clusters, the globular clusters in the local
Universe were formed before significant star formation had occurred in their
host galaxies, so that their source material was not yet ``polluted'' by
recycled material from evolved stars. On the other hand, clusters being formed
today are formed from interstellar material that must have undergone at least
a few cycles of stellar evolution (i.e., star birth and death, including
stellar mass loss through stellar winds and supernova explosions, both of
which processes replenish the ISM; the time-scale arguments are based on the
lifetimes of the higher-mass stars compared to those of their host galaxies as
such).

The connection between young or intermediate-age star cluster systems, as in
M82 B, and old globular clusters lends support to the hierarchical galaxy
formation scenario. Old globular clusters were once thought to have been
formed at the time of, or before, galaxy formation, i.e., during the first
galaxy mergers and collisions. However, we have now shown that the evolved CLF
of the compact star clusters in M82 B most likely to survive for a Hubble time
will probably resemble the high-mass wing of the ``universal'' old globular
cluster systems in the local Universe. Proto-globular cluster formation thus
appears to be continuing until the present.

\section{What determines the observed cluster masses?}

The observed CLF, and its corresponding CMF, is the result of both the
original mass distribution of the young star clusters and the GMCs they
originated from, and subsequent cluster disruption processes (e.g., Harris \&
Pudritz 1994, Bik et al. 2003, Boutloukos \& Lamers 2003, de Grijs et al.
2003a,c)

The apparent formation rate of the observed clusters in M82 B, i.e., the
number of clusters per unit age range, shows a gradual increase to younger
ages, reaching a maximum at the present time (see Fig. 4). This shows that
cluster disruption must have removed a large fraction of the older clusters
from the observed CLF (over and above the expected effects of evolutionary
fading of the stellar population, shown by the dotted line), assuming that the
cluster formation rate has been roughly constant over the lifetime of the
galaxy (but see below).

\begin{figure}
\psfig{figure=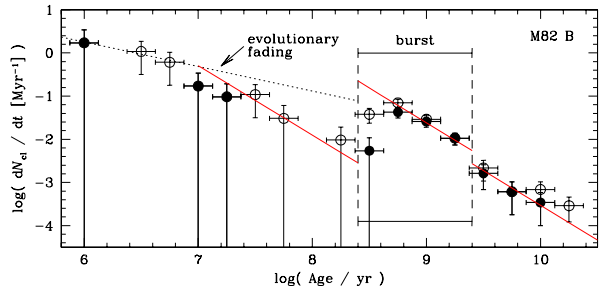,width=8.5cm}
\vspace{-4cm}
\caption[]{The cluster formation rate in M82 B (in number of clusters per Myr)
as a function of age. Open circles: our full sample; filled circles: clusters
with the most accurate age determinations. The vertical dashed lines indicate
the age range dominated by the burst of cluster formation. The dotted line is
the least-squares power-law fit to the fading, non-disrupted clusters, for a
constant ongoing cluster formation rate. The red solid line segments are the
disruption lines of clusters formed in the pre-burst phase, during the burst,
and in the post-burst phase, with a slope determined by that in the pre-burst
phase and shifted vertically to match the observed data points in the other 
age ranges.}
\end{figure}

The (initial) masses of the young clusters in M82 B brighter than our
observational completeness limit are mostly in the range $10^4 - 10^6 {\rm
M}_\odot$, with a median mass of $10^5 {\rm M}_\odot$ (de Grijs et al. 2001,
2003b,c). If the initial mass spectrum (i.e., the CMF) of the clusters formed
at the burst epoch was a power law (as expected from observations of very
young star cluster systems), disruption effects have transformed it into a
broader and flatter distribution on time-scales up to the system's present
lifetime, of $\lesssim 1$ Gyr. This has, in fact, been predicted by, e.g.,
Fall \& Zhang (2001) based on theoretical cluster disruption models.

Adopting a mass-dependent cluster disruption time-scale, $t_{\rm dis}$, in M82
B (i.e., $t_{\rm dis} \propto M_{\rm cl}^\gamma$, with $\gamma$ being the
power-law coefficient), and assuming instantaneous cluster disruption
(Boutloukos \& Lamers 2003; we realise that this is a physically unrealistic
assumption, but a more detailed, time-dependent treatment of cluster
disruption shows that this is in fact a reasonable approximation), we find
that the ratios between the real cluster formation rates in the pre-burst
phase ($\log[t/{\rm yr}]\ge 9.4$), the burst phase ($8.4 < \log[t/{\rm yr}] <
9.4$) and the post-burst phase ($\log[t/{\rm yr}] \le 8.4$) are roughly
$1:2:{1 \over 40}$ (de Grijs et al. 2003c). The formation rate during the
burst may have been higher if the actual duration of the burst was shorter
than adopted: uncertainties in the age determinations based on broad-band
colours may have broadened the peak (de Grijs et al. 2001, 2003b,c). The
cluster formation rate in the post-burst phase is much lower than in the
pre-burst phase because the burst has consumed a large fraction of the
available GMCs, leaving little material for cluster formation in the
post-burst phase (a similar effect is seen in NGC 1569, see below).

For the young star cluster systems in NGC 3310 and NGC 6745, for which their
respective characteristic cluster disruption time-scales (defined as the
mass-dependent disruption time-scale for a $10^4 {\rm M}_\odot$ cluster) are
significantly longer than the age of the starbursts in which they were formed,
the application of the empirical cluster disruption models of Boutloukos \&
Lamers (2003) results in an independent estimate of the {\it initial} cluster
mass function slope, $\alpha$. For the NGC 3310 and NGC 6745 cluster systems,
this slope $\alpha = 2.04(\pm 0.23)^{+0.13}_{-0.43}$ and $1.96 (\pm 0.15)\pm
0.19$, respectively, for masses $M_{\rm cl} \gtrsim 10^5 {\rm M}_\odot$ and
$M_{\rm cl} \gtrsim 4 \times 10^5 {\rm M}_\odot$ (de Grijs et al. 2003a).
Here, the first uncertainty represents the formal uncertainty in the fit and
the second the likely uncertainty resulting from our choice of fitting range.
These mass function slopes are consistent with those of other young star
cluster systems in interacting and starburst galaxies, circumnuclear starburst
rings, and of the H{\sc ii} region luminosity (and mass) functions in
``normal'' spiral and irregular galaxies (see de Grijs et al. 2003a for an
overview). They are also very similar to the mass function slopes of Galactic
open clusters and (OB) associations, of GMCs and their cores (although recent
observational evidence seems to indicate that the GMC mass distribution in
some galaxies in the local Universe, in particular in M33, may be
significantly deviating from that in the Milky Way and M31; cf. Engargiola et
al. 2003), and of the young compact clusters in the Large Magellanic Cloud
(Elson \& Fall 1985, Elmegreen \& Efremov 1997, Whitmore et al. 2002, Bik et
al. 2003).

Although it now appears that in the strong tidal fields of interacting
galaxies, and in the larger quiescent spirals and ellipticals, the process of
cluster formation and their resulting mass distribution may be comparable,
with $\alpha \simeq 2$, the interplay between the tidal field and gas supply
in the smaller starburst dwarf galaxies is clearly more complicated. For
instance, the young star cluster system in the nearby dwarf starburst galaxy
NGC 1569, mostly formed in the last $\sim 25$ Myr (Anders et al. 2003),
consists of two (possibly three; Hunter et al. 2000) well-known, massive SSCs,
in addition to a large number of lower-mass objects more similar in their
properties to Galactic open clusters. These objects may not survive for any
significant amount of time. This is supported by the observed change in the
CMF as the (current) burst proceeds (Anders et al. 2003), which is in the
sense expected for slowly dispersing open cluster-type objects. It is
therefore likely that the initial burst of cluster formation in NGC 1569
formed the massive SSCs, thereby depleting the gas supply for continuing
massive cluster formation -- the most recently formed objects are without
exception of open cluster or OB association appearance.

In summary, in this review I have shown that young, massive star clusters are
the most significant end products of violent star-forming episodes
(starbursts) triggered by galaxy collisions and gravitational interactions in
general. Their contribution to the total luminosity induced by such extreme
conditions dominates, by far, the overall energy output due to the
gravitationally-induced star-formation. The general characteristics of these
newly-formed clusters (such as their ages, masses, luminosities, and sizes)
suggest that at least a fraction may eventually evolve into equal, or perhaps
slightly more massive, counterparts of the abundant old globular cluster
systems in the local Universe, although they will likely be more metal rich
than the present generation of globular clusters. Establishing whether or not
such an evolutionary connection exists requires our detailed knowledge of not
only the physics underlying the evolution of ``simple'' stellar populations
(i.e., idealised model clusters), but also that of cluster disruption in the
time-dependent gravitational potentials of interacting galaxies. Initial
results seem to indicate that proto-globular clusters do indeed continue to
form today, which would support hierarchical galaxy formation scenarios.
Settling this issue conclusively will have far-reaching consequences for our
understanding of the process of galaxy formation and assembly, and of star
formation itself, both of which processes are as yet poorly understood.

\vfill\eject

\section*{Globular clusters}

{\it (Inset in a separate box)}
\vspace{0.5cm}

\noindent
The globular clusters in the Milky Way and most other large nearby spiral and
elliptical galaxies are thought to represent a fossil record of the conditions
at the time of galaxy formation. Their large ages, combined with their
generally low metal content, suggest that they were formed from largely
unpolluted -- pristine -- material, and therefore before significant star
formation and stellar evolution had taken place in their host galaxies. They
are thus often assumed to be the oldest galactic building blocks.

Globular clusters play a key role in astronomy: starting with John Herschel in
the 1830s, later followed by Shapley in 1917, the Milky Way globulars have
been instrumental in mapping the three-dimensional structure of our Galaxy.
Using variable stars in globular clusters, Shapley derived the distance to the
Galactic Centre for the first time. Although he based his distance estimate on
the incorrect assumption that the variable stars he detected were Cepheids
(while in fact they were RR Lyrae stars), thus leading to an overestimate of
the Galactic Centre distance by a factor of $\sim 2$, his work was of key
importance to our understanding that the Galactic Centre is far away from the
Sun. In fact, ever since globular clusters have been used extensively as the
basis for the extragalactic distance scale, by either using features based on
resolved stars (such as the ``turn-off magnitude'' in the observational
Hertzsprung-Russell diagramme, RR Lyrae variables, or the white dwarf cooling
sequence) or using the integrated globular cluster properties as a whole
(e.g., by using the globular cluster luminosity function as a ``standard
candle'', although its use -- and reliability -- is still controversial at
present).

From an astrophysical point of view, globular clusters constitute unique
laboratories for the study of stellar evolution and of the dynamics of dense
stellar systems: they are essentially single-age (i.e., almost instantaneously
formed), single-metallicity ``simple'' stellar populations of several $\times
10^5$ stars, thus providing statistically useful samples of even relatively
rare objects (such as blue stragglers, various types of variable stars,
planetary nebulae, white dwarfs, neutron stars and other X-ray sources, among
others). Furthermore, dynamical processes such as two or multiple-body
relaxation, core collapse, and mass segregation can be traced very robustly
using the resolved stelalr content of globular clusters in the Milky Way and
our nearest galactic neighbours.

\end{document}